\documentclass[a4paper,11pt]{article}
\pdfoutput=1 % if your are submitting a pdflatex (i.e. if you have
\usepackage{jcappub} % for details on the use of the package, please
\usepackage[T1]{fontenc} % if needed

\usepackage{comment}
\usepackage{caption}
\usepackage{subcaption}
\usepackage{graphicx} % needed for including graphics e.g. EPS, PS
\usepackage{pdfpages} % needed for including multi-page pdf documents
\usepackage{amssymb}
\usepackage{url}
\usepackage{appendix}
\usepackage{booktabs}

\usepackage{comment}
\usepackage{float}

\title{\bf The Bullet Cluster is not a Cosmological Anomaly}

\author{Craig Lage}
\author{and Glennys R. Farrar}
\affiliation{Center For Cosmology and Particle Physics\\ 
Department of Physics\\
New York University, New York, NY 10003, USA}

\emailAdd{csl336@nyu.edu}
\emailAdd{gf25@nyu.edu}

\abstract{The Bullet Cluster (1E0657-56) merger is of exceptional interest for testing the standard cold-dark-matter plus cosmological constant cosmological model, and for investigating the possible existence of a long- or short-range ``fifth-force'' in the dark sector and possible need for modifications of general relativity or even of Newtonian gravity.  The most recent previous simulations of the Bullet Cluster merger required an initial infall velocity far in excess of what would be expected within the standard cosmological model, at least in the absence of additional forces or modifications to gravity.   We have recently carried out much more detailed simulations than previously had been done, making pixel-by-pixel fits to 2D data-maps of the mass distribution and X-ray emission, allowing for triaxial initial configurations and including MHD and cooling.  Here, we compare the initial conditions of the Bullet Cluster merger as found in our new simulations to the initial conditions in similar-mass merging clusters in the Horizon cosmological simulation.  We conclude that our initial infall velocity, 2900 km/s at a separation of 2.5 Mpc, is consistent with $\rm \Lambda CDM$, given the inferred main cluster mass of $\rm 2 \times 10^{15} \, M_\odot$.  The initial concentration and shape found for the smaller (Bullet) cluster are typical for clusters of this mass range, but both quantities seem somewhat low for the larger (Main) cluster.  Due to the paucity of examples of clusters with such a high mass in simulations, these features of the main cluster cannot presently be used to test $\rm \Lambda CDM$.}

\begin{document}
\maketitle
%\flushbottom

\section{Introduction}

The ``Bullet Cluster'' is well-known and widely cited as a poster-child for Dark Matter (DM).  The observations show a clear separation between the 2D-projected mass distribution of gas revealed by the X-ray emission and the projected total mass revealed by gravitational lensing.  Interpreted in the standard cosmology with Dark Matter, this separation is a natural consequence of the shock-slowing of gas while the collisionless DM is unimpeded.   At the simplest level, this detachment of normal matter from the locus of gravitational lensing seems to rule out Modified Newtonian Dynamics (MOND) \cite{Milgrom_2008} as the explanation of Galactic rotation curves, but this conclusion has been contested \cite{Moffat}.  
A second means to test conventional physics with the Bullet Cluster, is to ask how rare its initial conditions are, in $\rm \Lambda CDM$ simulations.  Several simulation studies of the Bullet Cluster collision have already been performed.  Lee and Komatsu \cite{Lee_Komatsu} reviewed the simulations that had been done as of 2010, and found that, while the studies of Milosavljevic et al (\cite{milo2007}, hereafter M+07), and Springel and Farrar (\cite{sf07}, hereafter SF07) were consistent with a $\rm \Lambda CDM$ cosmology, the study of Mastropietro and Burkert (\cite{Mastropietro}, hereafter MB08) was not.  Lee and Komatsu estimated that the large initial infall velocity of the two clusters seen in MB08 had a probability of between $\rm 3.6 \times 10^{-9}$ and $\rm 3.3 \times 10^{-11}$ of occurring in a $\rm \Lambda CDM$ universe.  

Partially motivated by a desire to add clarity to this situation, we performed a detailed simulation of the Bullet Cluster where we compared the simulation to the observational data on a pixel-by-pixel basis in order to better constrain the initial conditions of the cluster collision.  This simulation study included triaxial initial clusters, magnetohydrodynamics, plasma cooling, and adaptive mesh refinement, and is reported on in detail in Lage and Farrar 2014 (\cite{LageFarrar2014} - hereafter LF14).  Our simulations in LF14 led to a best fit set of initial conditions with a somewhat lower initial velocity and a significantly larger mass for the combined clusters, as compared to the MB08 study.  As reported below, because of these differences the initial infall velocity of the Bullet Cluster collision deviates from the mean of $\rm \Lambda CDM$ N-body simulations by less than two standard deviations.  We also examine other aspects of the initial conditions, including concentrations, triaxiality ratios and impact parameter, and find no major inconsistencies with observations and N-body simulations.

The paper is organized as follows.  We begin by placing all of the existing simulations on a common footing with respect to initial infall velocity at a fixed separation.  Then we identify analog clusters in a large N-body simulation of the growth of cosmological structure, in order to quantitatively answer the question of whether the initial conditions of the Bullet-Cluster-merger precursor clusters are consistent with $\rm \Lambda CDM$.  Finally, we compare the cluster concentrations to observations and the shapes of the initial clusters to N-body simulations, and conclude.

\section{Review of Simulation Studies}

\begin{figure}[H]
%\begin{subfigure}[b]{\textwidth}
  \begin{subfigure}[b]{0.331\textwidth}
	\includegraphics[trim = 1.0in 0.0in 1.5in 0.34in, clip, width=\textwidth]{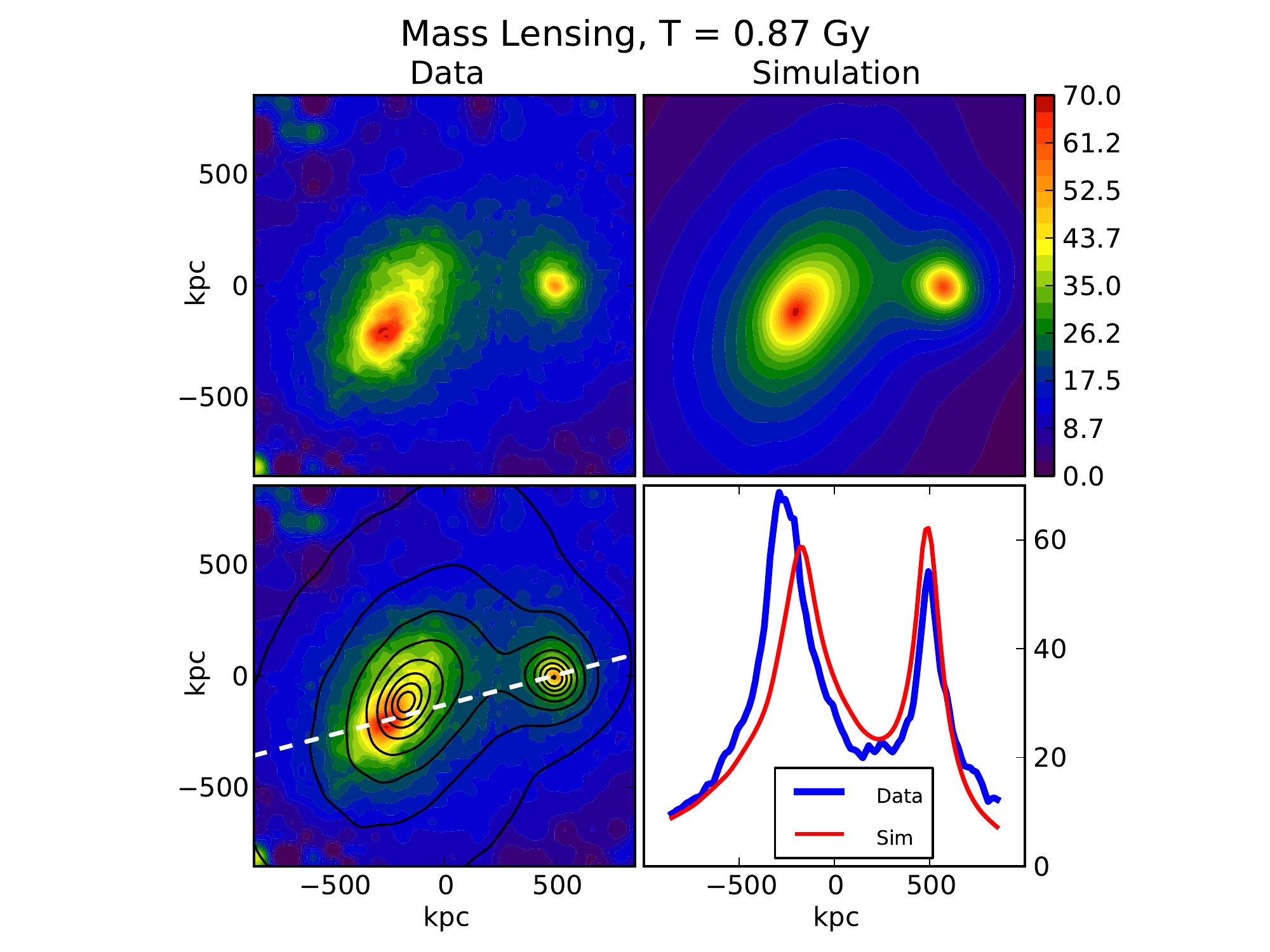}
	\caption{LF14 - Mass}
  \end{subfigure}
  \begin{subfigure}[b]{0.296\textwidth}
        \includegraphics[trim = 1.58in 0.0in 1.5in 0.34in, clip, width=\textwidth]{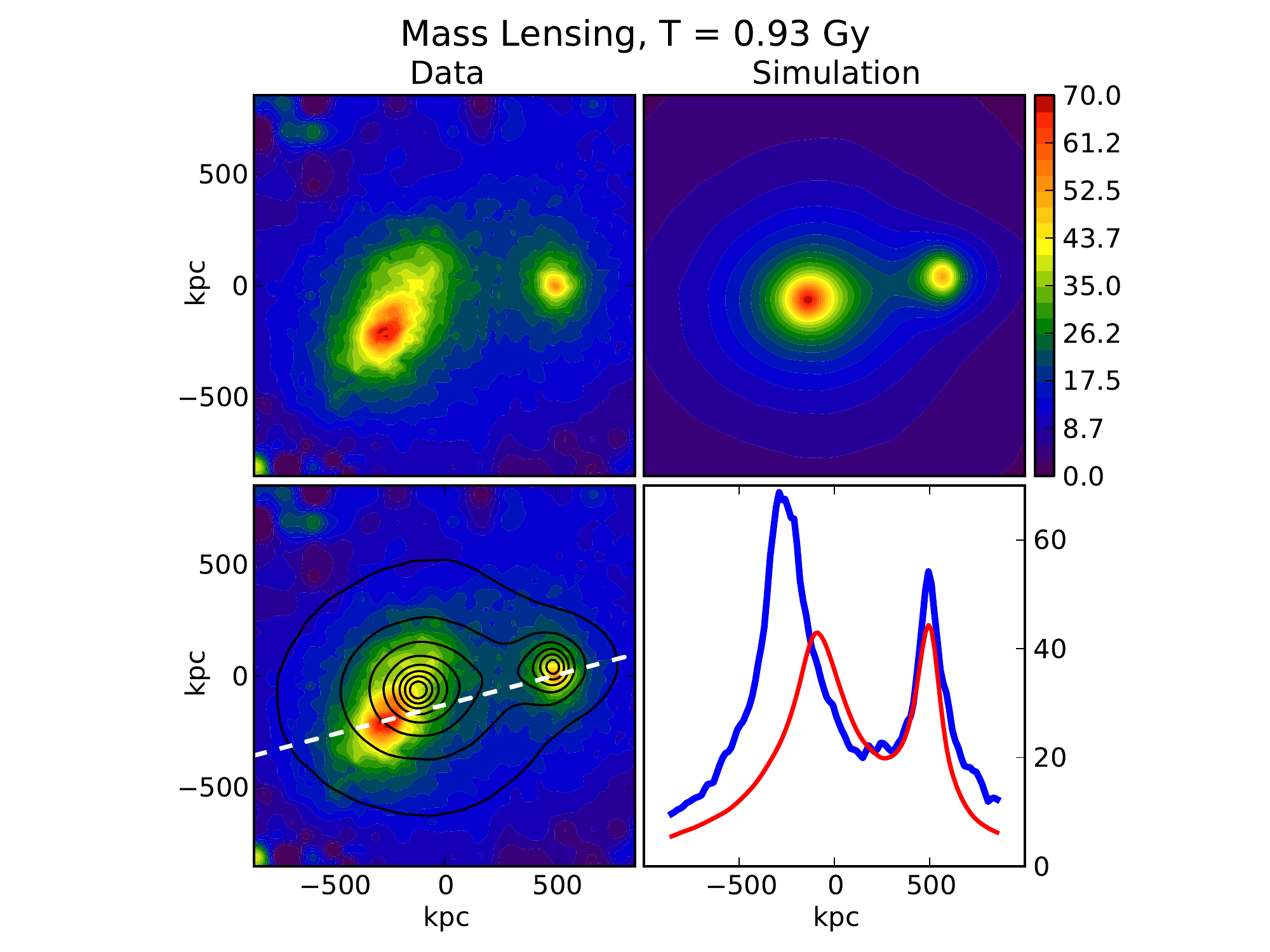}
	  \caption{SF07 - Mass}
  \end{subfigure}
  \begin{subfigure}[b]{0.344\textwidth}
    \includegraphics[trim = 1.58in 0.0in 0.7in 0.34in, clip, width=\textwidth]{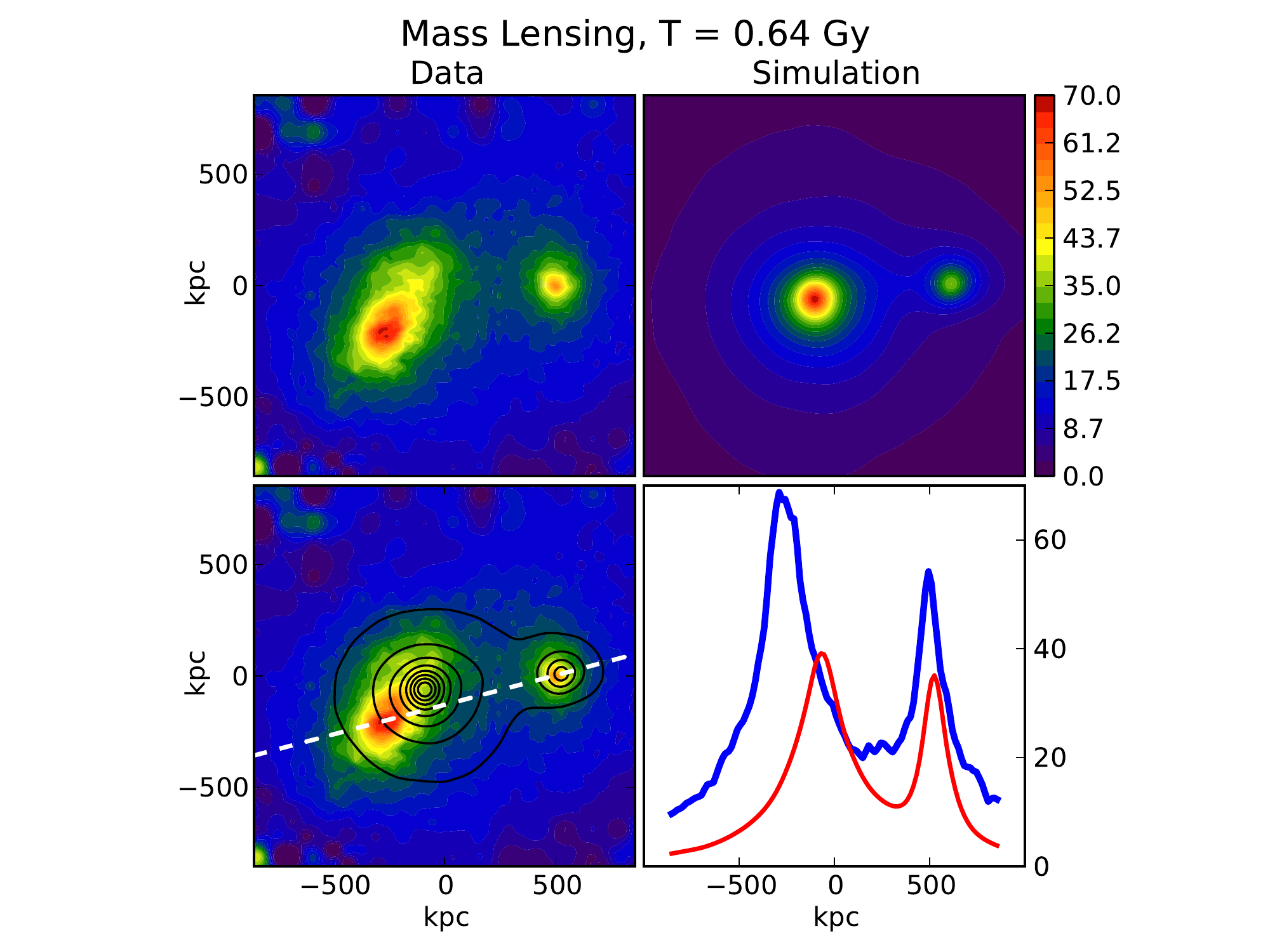}
    \caption{MB08 - Mass}
  \end{subfigure}

  \begin{subfigure}[b]{0.331\textwidth}
	\includegraphics[trim = 1.0in 0.0in 1.5in 0.34in, clip, width=\textwidth]{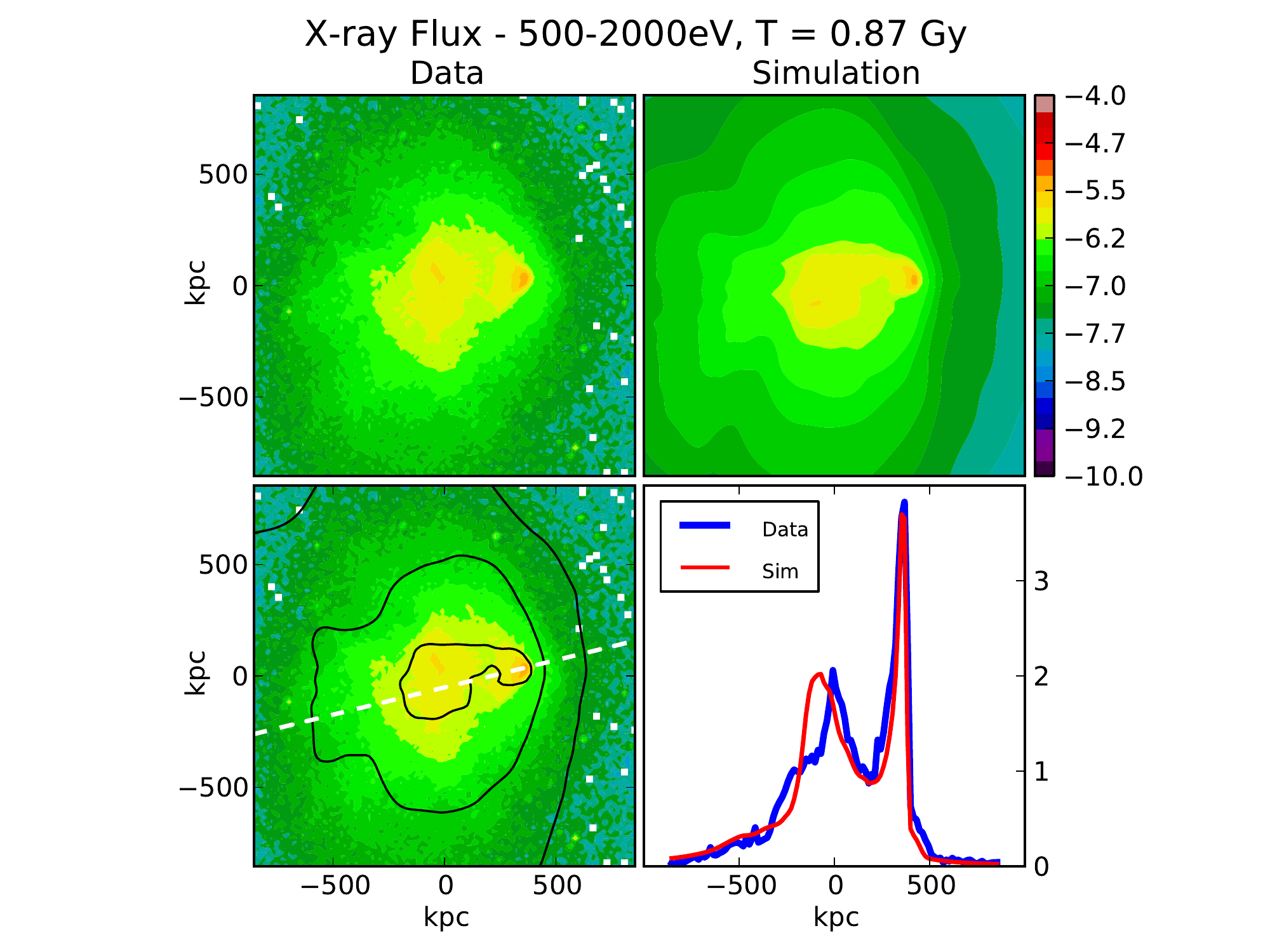}
	\caption{LF14 - X-ray}
  \end{subfigure}
  \begin{subfigure}[b]{0.296\textwidth}
        \includegraphics[trim = 1.58in 0.0in 1.5in 0.34in, clip, width=\textwidth]{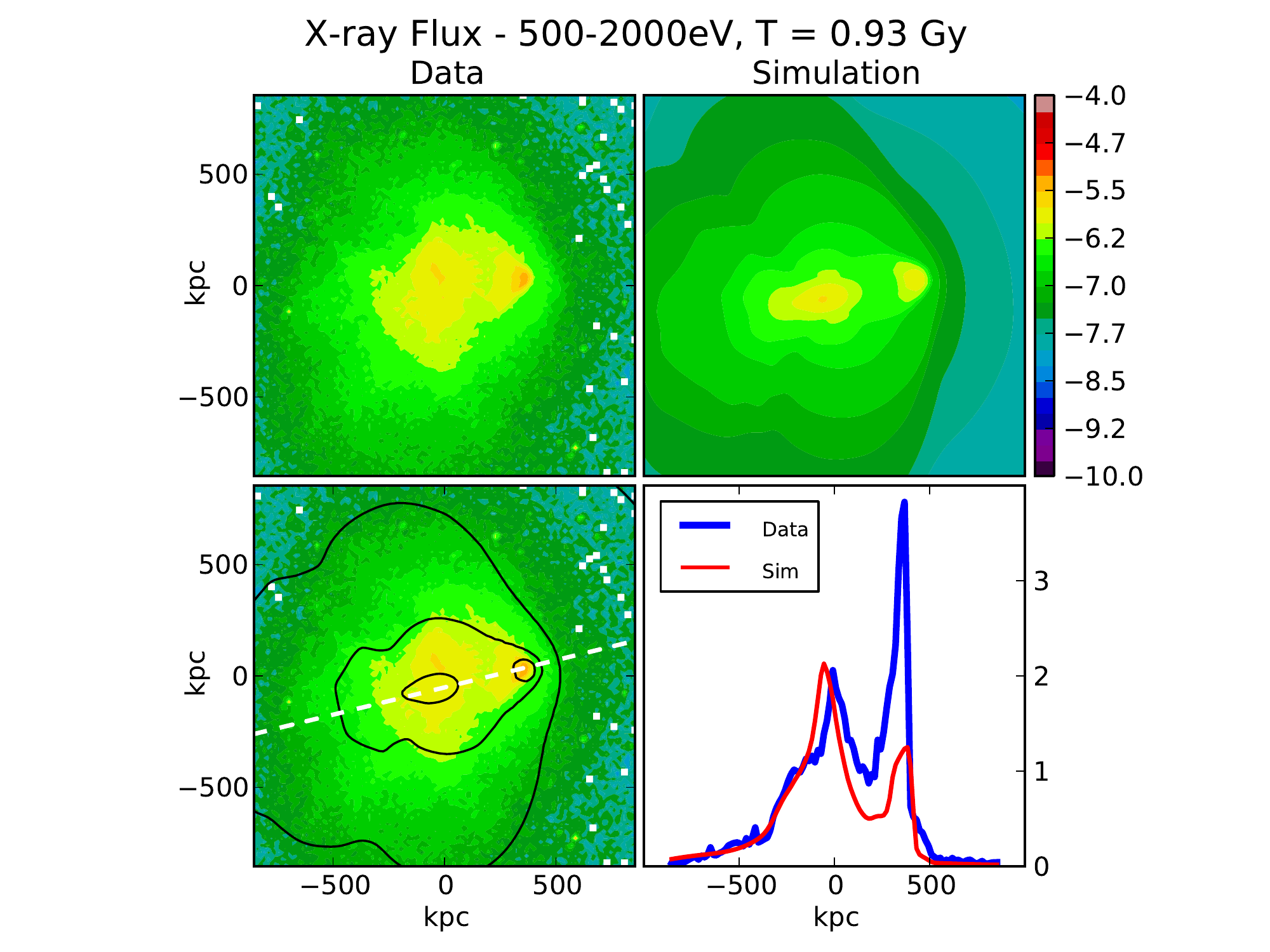}
	  \caption{SF07 - X-ray}
  \end{subfigure}
  \begin{subfigure}[b]{0.344\textwidth}
    \includegraphics[trim = 1.58in 0.0in 0.7in 0.34in, clip, width=\textwidth]{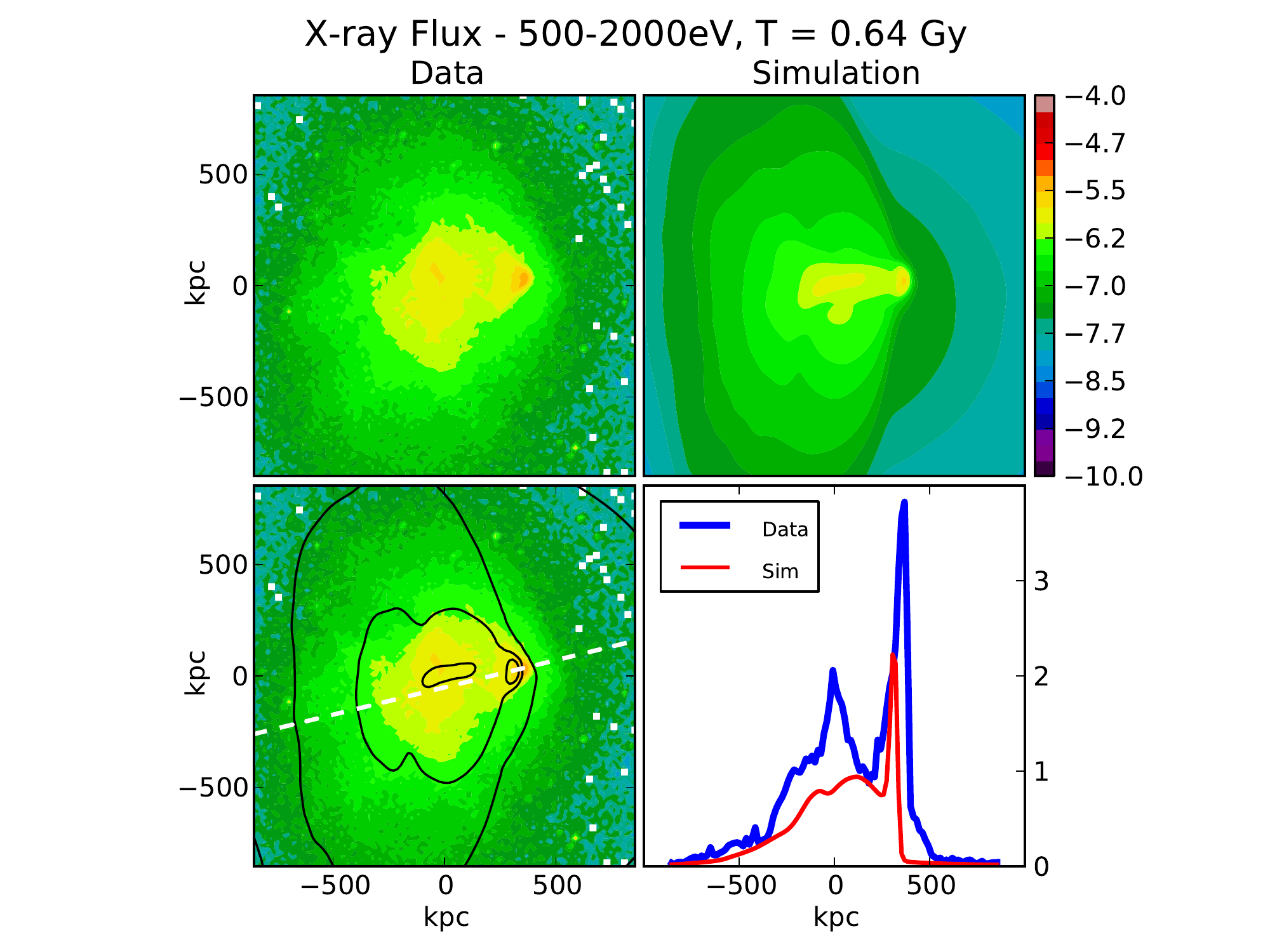}
    \caption{MB08 - X-ray}
  \end{subfigure}
	\caption{Mass lensing and 0.5-2.0 keV X-ray data as compared to best fits of LF14, SF07, and MB08.}
	\label{Fit_Plots}
	% These are from our best case and mastropietro_burkert and springel_farrar
	% trim is L B R T
\end{figure}

The studies of SF07 and MB08 only attempted to constrain a small number of extracted parameters, specifically the spacings between the mass centroids and the X-ray peaks, and to qualitatively resemble the general shape of the X-ray flux maps.   Our work in LF14 was a more detailed study which minimized a $\chi^2$ figure of merit between the two-dimensional observational data sets and the simulation.  (The $\chi^2$ parameter is used only as a figure of merit to optimize and compare quality of different fits;  since the $\sigma$ in the denominator does not include astrophysical noise such as expected from small scale inhomogeneities, even an excellent fit will not have $\chi^2 = 1$.)  This work included triaxial initial clusters, magnetohydrodynamics, plasma cooling, and adaptive mesh refinement.  More than 1000 sets of initial conditions were tested in order to find the best fit, with the result that the simulation in LF14 fits cluster observations significantly better than past simulations.  Figures \ref{Fit_Plots} shows the improved fit we achieved in LF14 as compared to SF07 and MB08.

The figure of merit $\rm \chi^2$ values are calculated by comparing the mass lensing and lowest energy X-ray data to the simulations as described in LF14. The resulting $\rm \chi^2$ values are 3.92 for LF14, 13.67 for SF07, and 19.93 for MB08.  The greatly improved fit motivates us to revisit the question of consistency with $\rm \Lambda CDM$, within the framework of the LF14 best fit simulation. 

Table \ref{Simulation_Comparisons} shows a comparison of the cluster masses, initial cluster separations and initial infall velocities found in the various studies. (For brevity, in what follows we refer to the larger cluster as the main cluster, and the smaller cluster as the subcluster.)  To facilitate comparison, we also give a standardized initial infall velocity calculated assuming that the clusters move as point masses on a ballistic trajectory from their starting separation to a separation of 2500 kpc; since there is very little interaction between the clusters at separations larger than 2500 kpc, assuming a ballistic trajectory of these widely separated clusters is a very good approximation, as shown in Figure \ref{Trajectory_Comparison}.  We have included the simulation study of Milosavljevic \cite{milo2007} in Table \ref{Simulation_Comparisons}, although since it is a 2D axisymmetric simulation, it is not in the same category as the other studies.  Also included is another early study, the dark-matter-only simulation of Randall et al (\cite{Randall_2008}, hereafter R+08).   The initial velocity and separation of the Randall study were obtained from personal communication with the author; the masses are calculated from halo parameters given in R+08.  A more recent dark-matter-only study has been performed by Dawson \cite{Dawson2013}, but comparable initial condition information is not available; from Figures 2 and 4 of \cite{Dawson2013}, the results appear to be largely consistent with SF07 (W. Dawson, private communication).

\begin{table}[H]   
	\centering
	\begin{tabular}{lcccccc} 
		\hline 
		Authors & $\rm M_{Main}$ & $\rm M_{Sub}$ & $\rm R_{Initial}$ & $\rm V_{Initial}$ &  $\rm V_{2500}$ &$\rm \chi^2$\\ 
		 & $\rm (M_\odot)$ & $\rm (M_\odot)$ & (kpc) & (km/sec)  &  (km/sec)&\\ 
                \hline
		M+07 & $\rm 1.27 \times 10^{15}$ & $\rm 2.54 \times 10^{14}$ & 4600 & 0 & 1546 & --\\
                R+08 & $\rm 8.56 \times 10^{14}$ & $\rm 5.25 \times 10^{14}$ & 4000 & 4100 & 4225 & --\\
		SF07 & $\rm 1.50 \times 10^{15}$ & $\rm 1.50 \times 10^{14}$ & 3370 & 2057 & 2386 & 13.67\\
                MB08 & $\rm 7.13 \times 10^{14}$ & $\rm 1.14 \times 10^{14}$ & 5000 & 3000 & 3228 & 19.93\\ 
		LF14 & $\rm 1.91 \times 10^{15}$ & $\rm 2.59 \times 10^{14}$ & 2800 & 2799 & 2943 & 3.92\\

		\hline
	\end{tabular} 
	\caption{Comparison of initial infall velocities from different simulation studies.  The $\rm V_{2500}$ column gives a standardized initial infall velocity calculated assuming that the clusters move as point masses on a ballistic trajectory from their starting separation to a separation of 2500 kpc.  The calculation of the $\rm \chi^2$ parameter, which measures the fit between the simulations and the observations, is described in detail in LF14.}
	\label{Simulation_Comparisons}
\end{table}

\section{Cluster Initial Velocities}
\begin{figure}[H]
  \centering
  \includegraphics[trim=0.0in 0.0in 0.0in 0.0in,clip,width=0.60\textwidth]{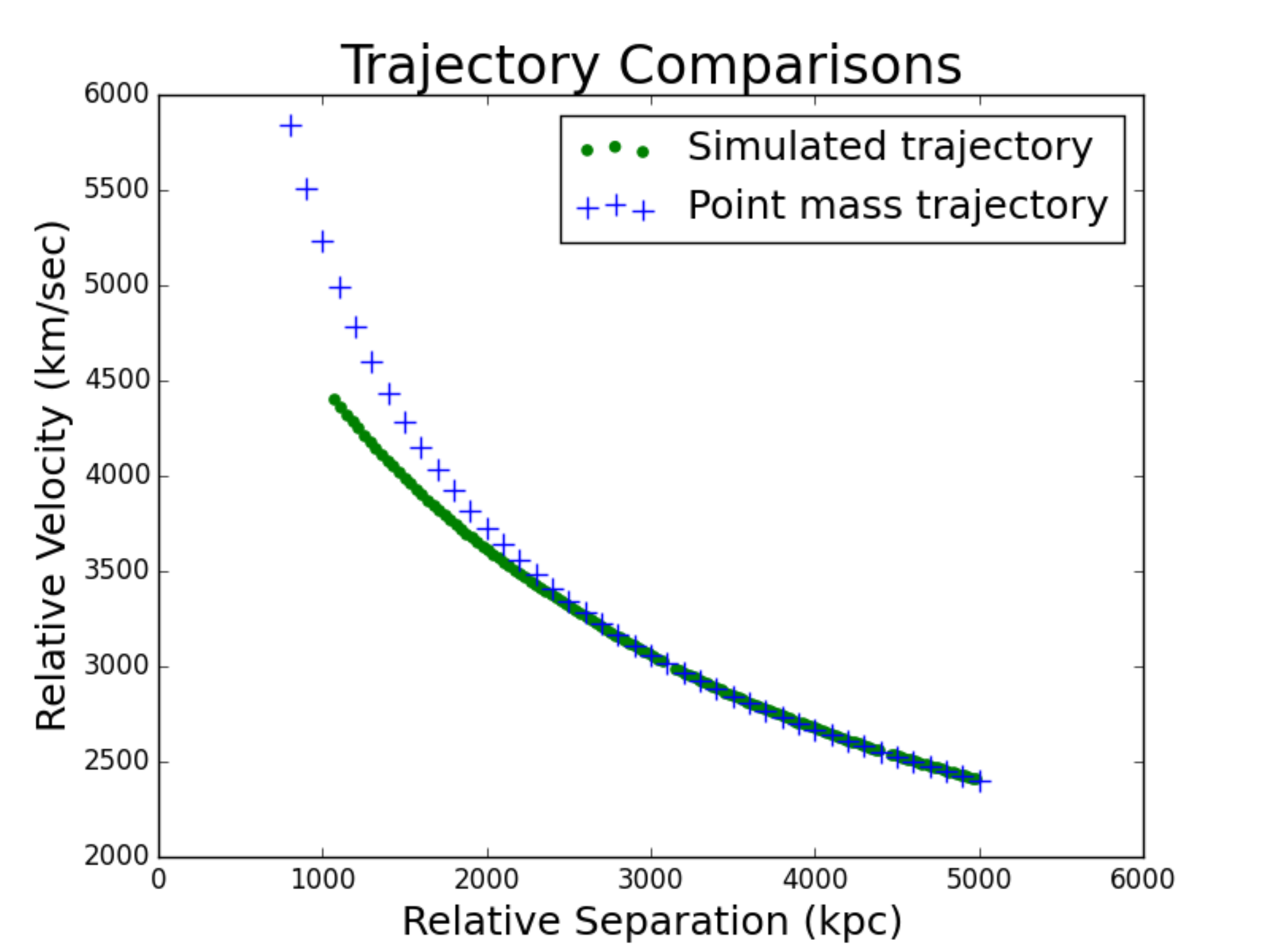}
  \caption{Comparison of simulated vs calculated trajectories.  The green circles show the full Enzo simulation of the clusters detailed in LF14, starting at an initial separation of 5000 kpc.  The blue crosses show the trajectory calculated assuming that the clusters are two point masses.  It is seen that the point mass assumption is relatively good down to a cluster separation of about 2500 kpc, when the virial radii begin to overlap.  Thus, it is valid to use the point mass assumption to normalize the different simulations.}
  \label{Trajectory_Comparison}
\end{figure}

To estimate whether the initial velocities of these simulations are consistent with a $\rm \Lambda CDM$ cosmology, we use an N-body simulation known as the Horizon Run (Kim \cite{Horizon_Paper}).  This is a large dark-matter-only simulation using $4120^3 = 6.99\times 10^{10}$ particles, and covering a volume of $\rm (6.59 Gpc/h)^3$.  We analyze the data from this simulation in the following manner: 
\begin{enumerate}
  \item We start with the database of halos from the $\rm z=0.5$ snapshot.  This database contains the masses, locations, and velocities of approximately 1.1 million halos.  The z=0 and z=0.5 snapshots were available to us, and we used the z=0.5 snapshot since it is close to the redshift at the beginning of the simulation, which is approximately z=0.39.
  \item For a range of target masses between $\rm 5 \times 10^{14} \, M_\odot$ and $\rm 2 \times 10^{15} \, M_\odot$, we search for a cluster within 10\% of the target mass.  A cluster meeting this criterion is referred to as a main cluster analog.
  \item For each of these ``main'' clusters, we search for a neighboring cluster separated from the main cluster by a distance between 2500 kpc and 5000 kpc, with a mass between 6 times and 10 times less than the main cluster analog.  A cluster meeting these criteria is referred to as a subcluster analog.
  \item We extract the relative velocities of each pair of clusters, and convert to the value at a separation of 2500 kpc, assuming that the clusters move as point masses along ballistic trajectories from their current separation to a separation of 2500 kpc.  Figure \ref{Trajectory_Comparison} shows that this is a valid assumption.
  \item We also extract the total energy and impact parameter of these two clusters.
\end{enumerate}

\begin{figure}[ht]
  \centering
  \includegraphics[trim=0.0in 0.0in 0.0in 0.0in,clip,width=0.60\textwidth]{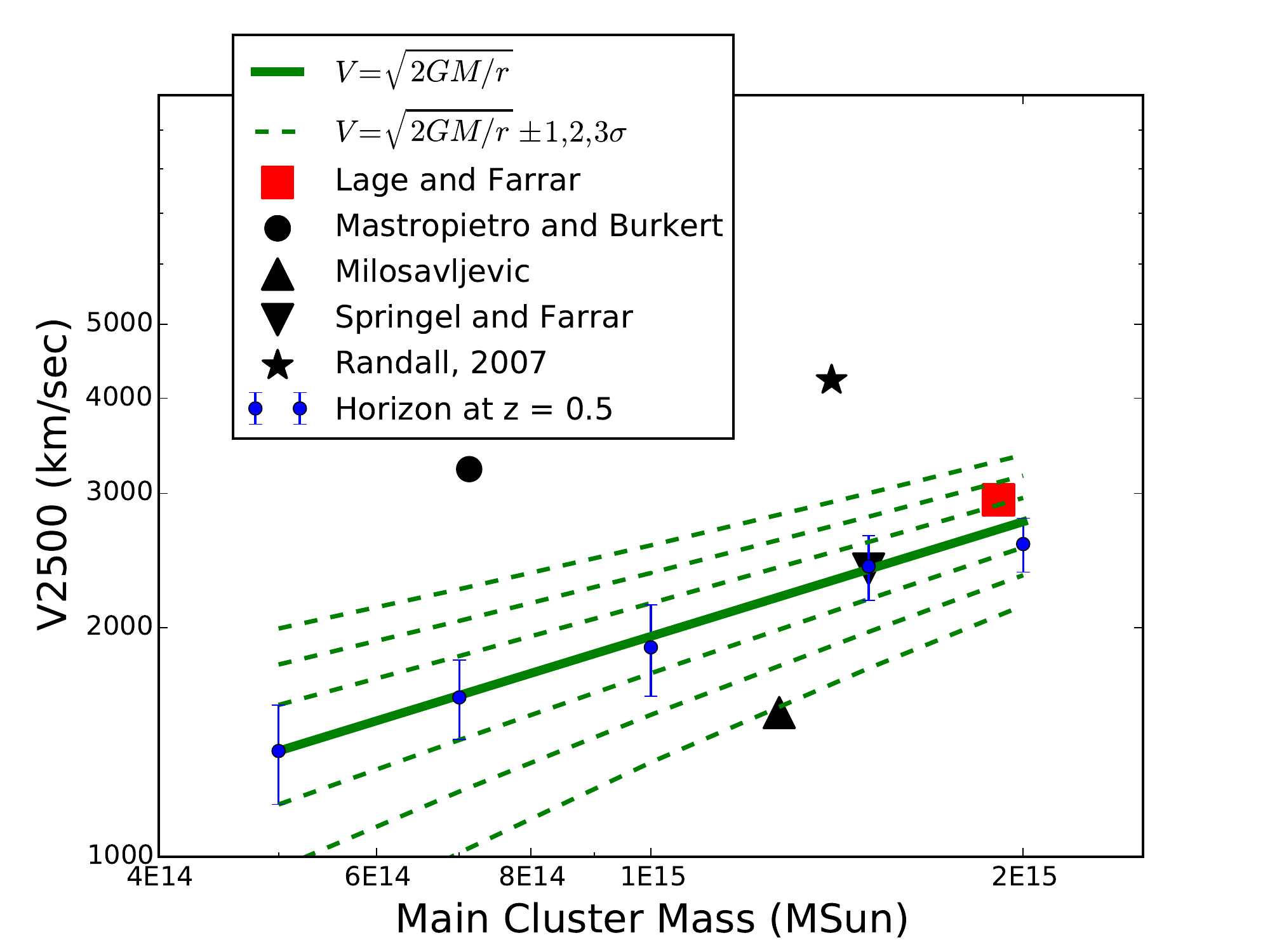}
  \caption{Initial infall velocity of the subcluster relative to the main cluster extracted from the z=0.5 snapshot of the Horizon simulation at a separation of 2500 kpc, using the analysis technique described in the text.  The small circles are the mean relative velocity, with 1 $\sigma$ error bars.  The thick solid line shows the expected $\rm V=\sqrt{2 G M_{Main}/R_{2500}}$ behavior.  The three dotted lines are the $\rm V=\sqrt{2 G M_{Main}/R_{2500}}$ curve offset by 1, 2, and 3 $\sigma$, respectively.}
  \label{Velocity_vs_Mass}
\end{figure}

Figure \ref{Velocity_vs_Mass} shows the initial infall velocities extracted in this way compared to the above simulation studies.  Since the main cluster mass is much larger than the subcluster mass, we expect the initial infall velocities to be equal to $\rm \sqrt{2 G M_{Main}/R_{2500}}$, and this is just what is seen in Figure \ref{Velocity_vs_Mass}.  The fit to the expected behavior improves at lower masses because there are many more clusters and hence less stochastic variability.  While there are 2204 cluster pairs whose main cluster mass is $\rm 5 \times 10^{14} \, M_\odot \, \pm 10\%$ there are only 4 cluster pairs at $\rm 2 \times 10^{15} \, M_\odot \, \pm 10\%$.  Because of the much larger number of cluster pairs at lower cluster masses, we use the mean and standard deviation calculated at a main cluster mass of $\rm 5 \times 10^{14} \, M_\odot$ and extrapolate to larger masses, rather than using the mean and standard deviation calculated at the larger masses.  

The parameters obtained in the simulation studies discussed above are also plotted in Figure \ref{Velocity_vs_Mass}.  The best fit initial infall velocity from LF14 is about 1.24 standard deviations above the mean of the $\rm \Lambda CDM$ distribution, for the mass as determined by LF14.

Although two of the earlier studies are 5 standard deviations and one is 3 standard deviations from the mean, greater reliance can be placed on the LF14 initial conditions, since the LF14 simulation and fitting method was superior to earlier efforts, as discussed above.  Therefore we conclude that the tension between the initial infall velocity of the Bullet Cluster and expectations from $\rm \Lambda CDM$ exposed by Lee and Komatsu [3] can be considered to be resolved. 

\begin{figure}[ht]
  \centering
  \includegraphics[trim=0.0in 0.0in 0.0in 0.0in,clip,width=0.50\textwidth]{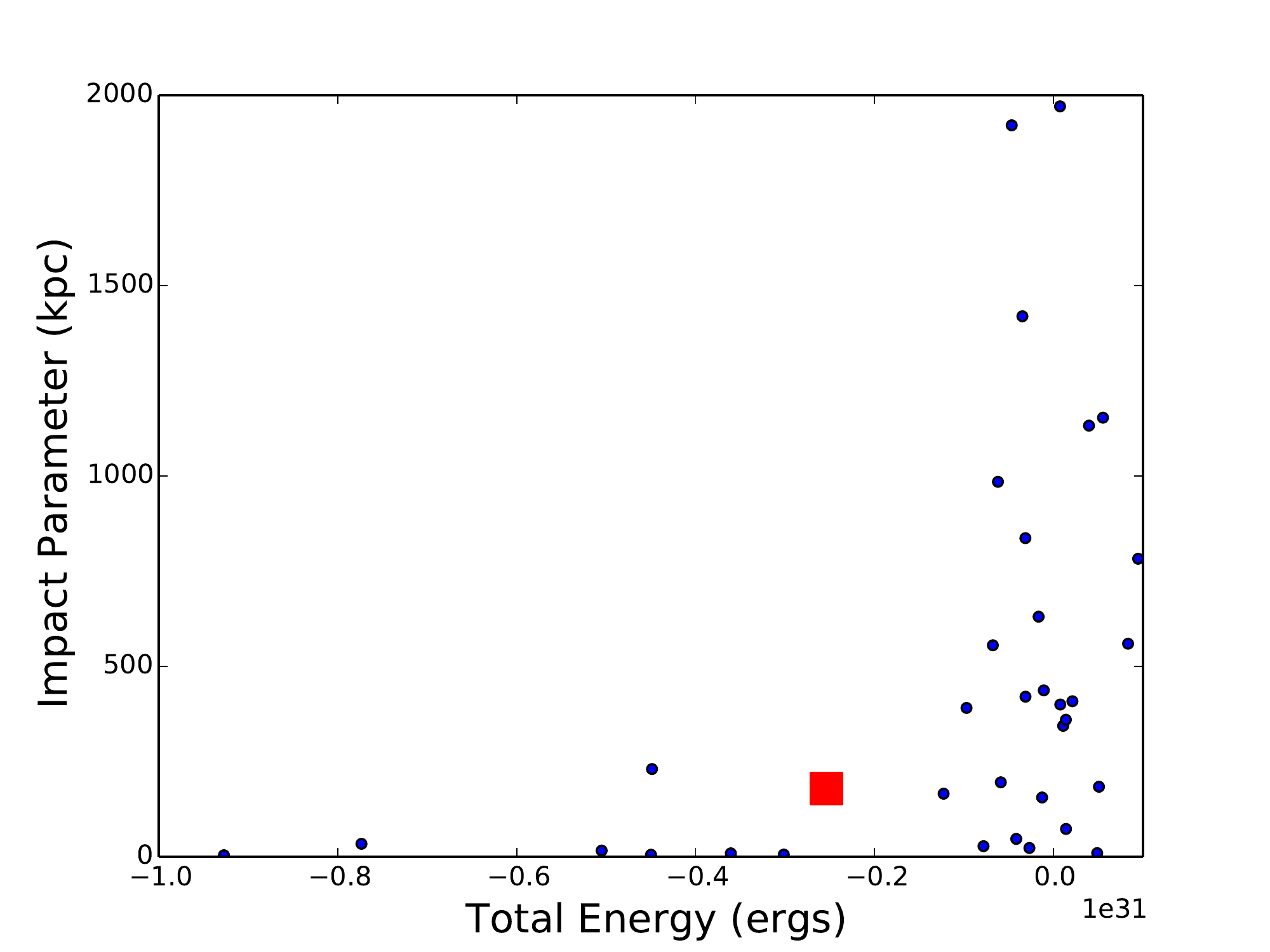}
  \caption{Total Energy vs Impact Parameter of cluster pairs in the Horizon Run having main cluster mass of $\rm 2 \times 10^{15} \, M_\odot \, \pm 30\%$.  The large square represents the best fit simulation from LF14.}
  \label{Energy_vs_IP}
\end{figure}

In Figure \ref{Energy_vs_IP}, we plot the total energy and impact parameter of pairs of clusters extracted as described above, with main cluster mass of $\rm 2 \times 10^{15} \, M_\odot \, \pm 30\%$, as compared to the LF14 best fit simulation.  It is seen that most cluster pairs in the Horizon simulation are near zero total energy, and the LF14 best fit simulation falls comfortably within the distribution.

\section{Cluster Concentrations and Shapes}
The concern that the Bullet Cluster is inconsistent with $\rm \Lambda CDM$ cosmology has focused on the initial infall velocity of the colliding clusters, and we have shown in the preceding section that this velocity is in fact not exceptional.  However, it is also worthwhile to examine the consistency of the sizes and shapes of the colliding clusters with observations and N-body simulations based on $\rm \Lambda CDM$.  

\begin{figure}[ht]
  \begin{subfigure}[b]{0.45\textwidth}
        \includegraphics[trim=3.0in 1.5in 3.0in 2.0in,clip,width=\textwidth]{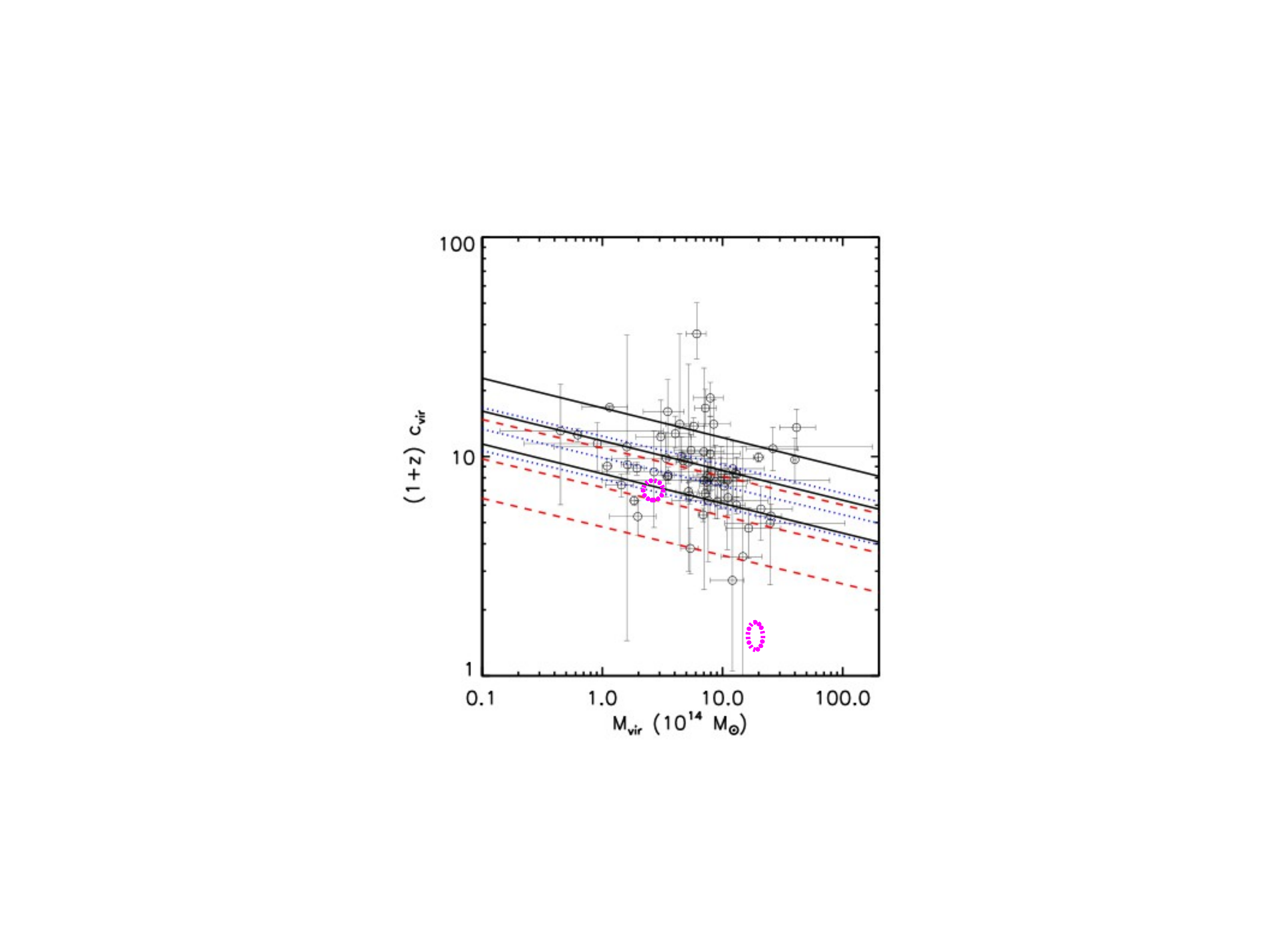}
        \caption{Figure reproduced from Comerford et.al. \cite{Comerford}.}
  \end{subfigure}
  \begin{subfigure}[b]{0.45\textwidth}
       \includegraphics[trim=3.0in 1.0in 3.0in 2.0in,clip,width=\textwidth]{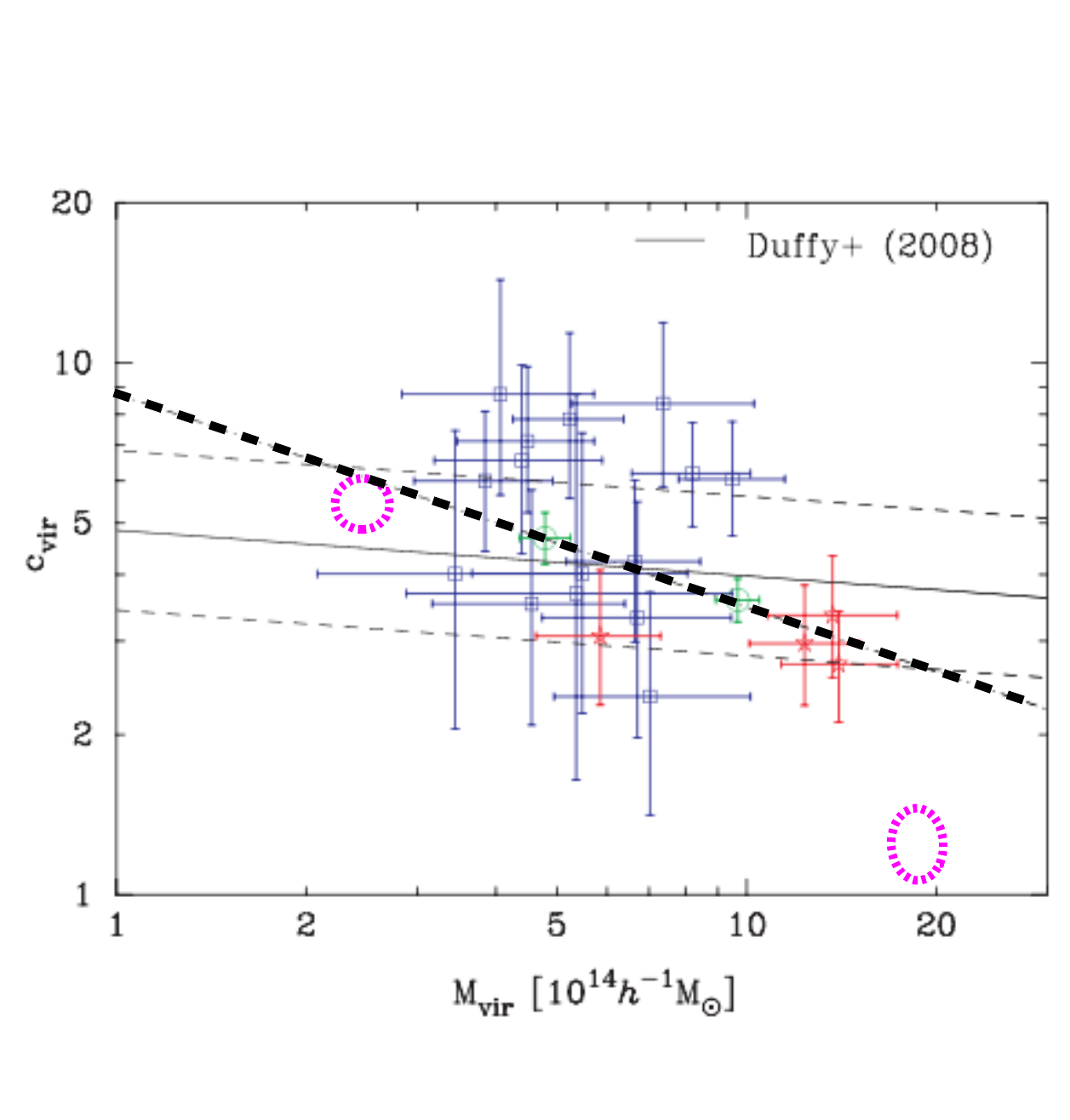}
	\caption{Figure reproduced from Okabe et.al. \cite{Okabe2010}.}
  \end{subfigure}
  \caption {Comparison of masses and concentrations from the LF14 best fit simulation to measured mass-concentration relations.  The dotted ellipses represent one-sigma errors around the LF14 best fit initial conditions.  In both plots, the subcluster is on the left and the main cluster on the right.}
  \label{Mass_Con}
  % Trim is Left Bottom Right Top
\end{figure}

First, we examine the concentrations of the colliding clusters and compare these to observations.   As used here, the concentration is defined as the ratio of the virial radius to the NFW scale radius as follows: $\rm c = R_{200} / R_s$.  Figure \ref{Mass_Con} shows the LF14 best fit masses and concentrations as compared to two observational studies.  Figure \ref{Mass_Con}(a) shows the comparison to the work of Comerford \cite{Comerford}.  While the subcluster is quite typical, the main cluster appears to have an unusually low concentration for its mass.  However, a more recent study of Okabe \cite{Okabe2010}, shown in Figure \ref{Mass_Con}(b), has found a steeper slope for the Mass-Concentration relationship (heavy dashed line in Figure \ref{Mass_Con}(b)) which is more consistent with our findings for the main cluster.  We emphasize that in the full simulation study detailed in LF14 a wide range of concentrations were explored, and the ones reported here are the best fit.  The low concentration of the main cluster found there was necessary in order to fit the observations.

In order to quantify the shape of these clusters, we introduce a set of axis ratios.  We assume that the clusters are triaxial ellipsoidal shapes, characterized by the lengths of each of the three axes of the ellipsoid.  The shape of the ellipsoid is then completely determined by the two ratios of these three axes, with P being the shortest axis to longest axis ratio, and Q being the intermediate axis to longest axis ratio.  With these definitions, we have $\rm 1.0 \geq Q \geq P \geq 0.0$.  If $\rm Q = 1$, then the ellipsoid is an oblate spheroid, with a shape like a pancake.  If $\rm P = Q$, then the ellipsoid is a prolate spheroid, with a shape like a cigar.  If all three axes are equal, then the shape is a sphere, and if all three axes are unequal, the shape is triaxial.  

\begin{figure}[ht]
  \begin{subfigure}[b]{0.45\textwidth}
        \includegraphics[trim=3.3in 1.5in 3.3in 1.5in,clip,width=\textwidth]{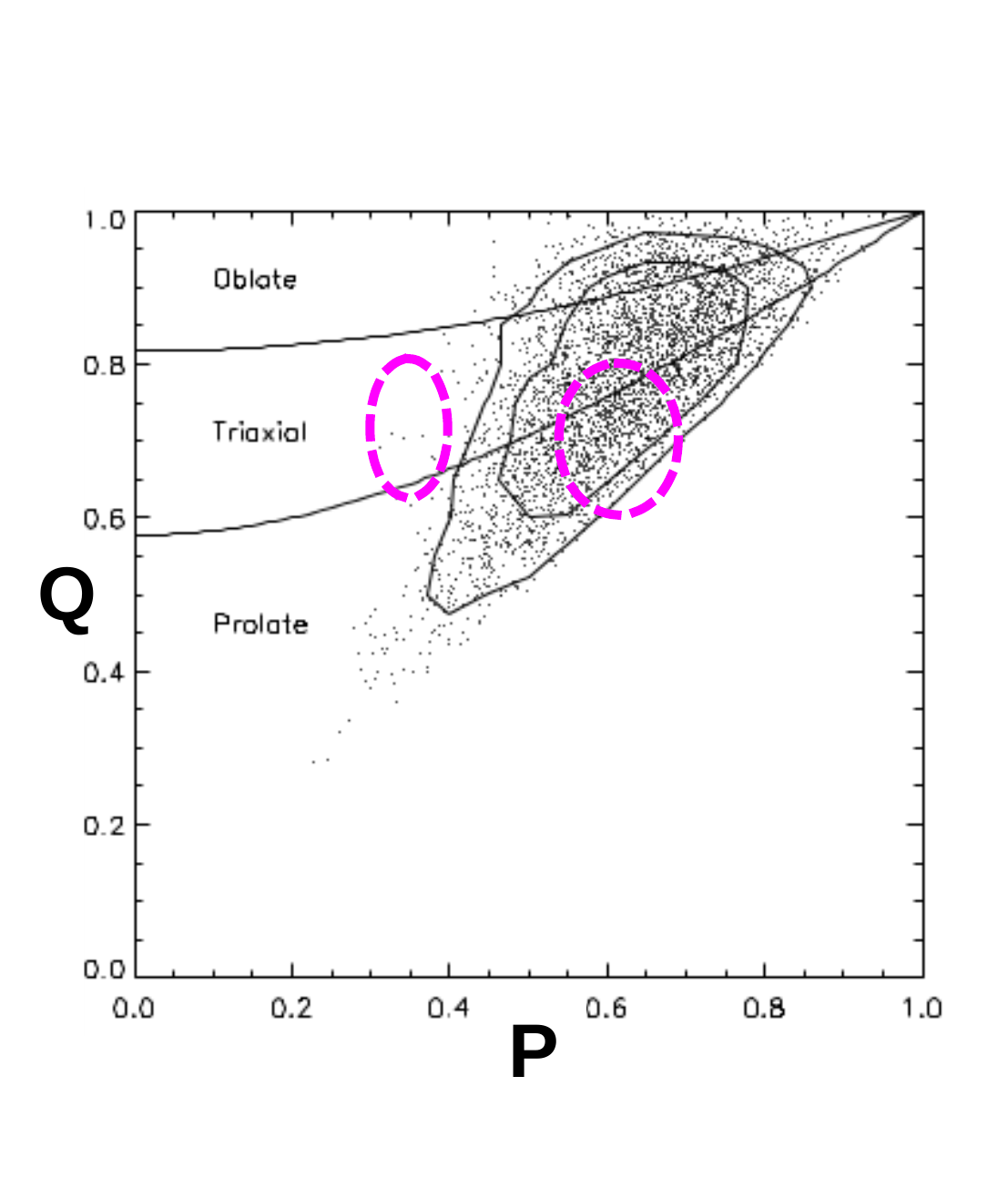}
        \caption{Axis ratios of halos extracted from N-body simulations by Bailin et.al.\cite{bailin2005}.  The dotted ellipses represent one-sigma errors around the LF14 best fit axis ratios, with the main cluster on the left and the subcluster on the right.}
  \end{subfigure}
  \hspace{5mm}
  \begin{subfigure}[b]{0.45\textwidth}
         \includegraphics[trim = 2.3in 1.2in 2.3in 1.2in, clip, width=\textwidth]{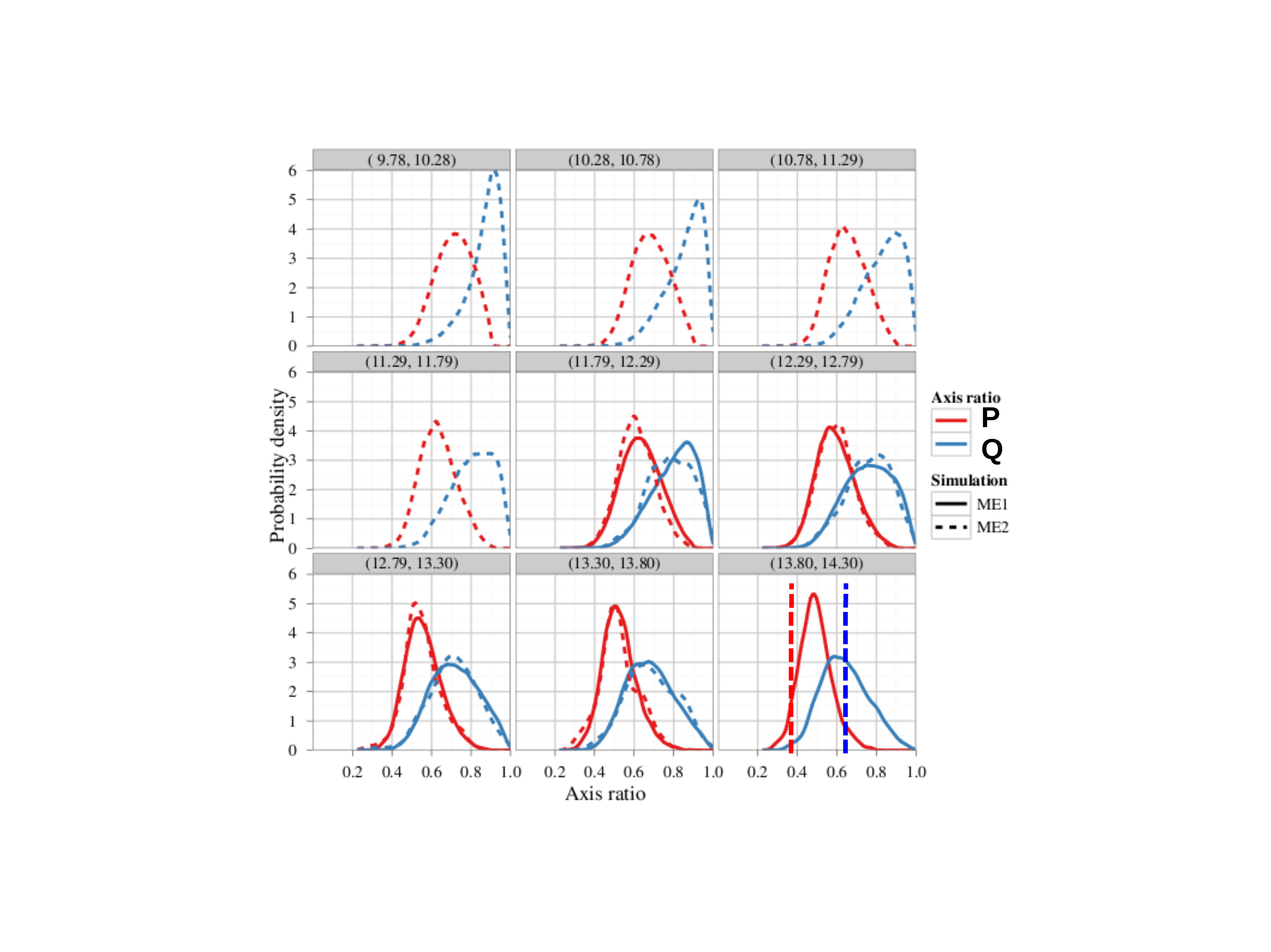}
	\caption{Axis ratios of halos of different masses extracted from N-body simulations by Schneider, Frenk, Cole \cite{Schneider2012}, showing that more massive clusters have smaller axis ratios.  The numbers in parentheses are the mass ranges in $\rm \log(M_\odot)$, with masses increasing from upper left to lower right.  The dotted lines in the lower right panel are the LF14 best fit axis ratios for the main cluster.}
  \end{subfigure}
  \caption {Comparison of the LF14 best fit axis ratios to those extracted from N-body simulations.}
  \label{Cluster_Shapes}
\end{figure}

Figure \ref{Cluster_Shapes}(a) compares the LF14 best fit axis ratios to an N-body simulation study by Bailin \cite{bailin2005}.  The subcluster is well within the population of clusters, while the small axis ratio of 0.35 found for the main cluster appears somewhat unusual.  The more detailed study of Schneider, Frenk, and Cole \cite{Schneider2012}, shown in Figure \ref{Cluster_Shapes}(b), examines the trends of axis ratios as a function of cluster mass and finds that more massive clusters tend to have smaller axis ratios, although the large mass of the main cluster (nearly $\rm 2\times 10^{15} \, M_\odot$) is actually beyond the range considered.  The lower right panel of Figure \ref{Cluster_Shapes}(b) shows the LF14 best fit axis ratio for the main cluster as compared to the largest masses studied.  While we are unable to quantify how likely the LF14 best fit axis ratio of 0.35 is, the trend of more massive clusters having smaller axis ratios is in the right direction.
%\clearpage
\section{Conclusions}
\label{Summary and Conclusions}

Ever since the initial report of a high relative velocity between the components of the merging Bullet Cluster, 4740 km/s at the time of observation \cite{markevitch05}, the possibility that the Bullet Cluster may require a new force between dark matter particles \cite{fr07}, or be incompatible with $\rm \Lambda CDM$ cosmology\cite{hwRareBullet06}, has been a topic of interest.  Ref. \cite{sf07} pointed out that the relative velocity estimated from the bow shock\cite{markevitch05} is not the relative velocity of the Dark Matter clusters, mitigating the issues raised in \cite{fr07,hwRareBullet06}.  However the more recent analysis by \cite{Lee_Komatsu} comparing the initial conditions reported in detailed merger simulations to configurations found in $\rm \Lambda CDM$ cosmological simulations, found a seemingly serious problem:  the best-fit initial conditions of \cite{Mastropietro} (MB08) -- then the most recent and in principle comprehensive attempt to simulate the merger -- had a probability of $3 \times 10^{-9} - 3 \times 10^{-11}$ of occurring in $\rm \Lambda CDM$.

We recently performed a detailed simulation of the Bullet Cluster using a new approach, fitting the simulation to the observational data on a pixel-by-pixel basis.  The simulation included triaxial initial clusters, magnetohydrodynamics, plasma cooling, and adaptive mesh refinement \cite{LageFarrar2014} (LF14).  Besides being more complete in terms of physical modeling than earlier simulations, the fit achieved to the mass-lensing and X-ray data by LF14 is significantly better than in previous simulations.  In the present work, we compare the initial infall velocities, impact parameters, and shapes of the intial clusters found in the LF14 fit to the Bullet Cluster observations, to the distributions found in the Horizon N-body simulation of structure growth in $\rm \Lambda CDM$ cosmology. 

Our most important result is that the initial infall velocity as determined by LF14 is entirely compatible with expectations from $\rm \Lambda CDM$ cosmology.   Since the LF14 simulation gives the (by far) best description of the observations \cite{LageFarrar2014}, the tension between the Bullet Cluster infall velocity and $\rm \Lambda CDM$ cosmology is now removed.   However this is not the only lesson to be learned.  Figure \ref{Velocity_vs_Mass} collects the initial masses and infall velocities determined by the various simulations of the Bullet Cluster merger, at a common initial separation of 2.5 Mpc, and compares them to the distribution for analog-clusters found in the Horizon simulation.   Most of the earlier simulations dramatically disagree with the $\rm \Lambda CDM$ expectations, with two being 5-sigma above the mean relationship and one being 3-sigma below.   This highly disparate behavior arises because the inferred initial conditions of the various merger simulations differ widely (c.f., Table 1). The mass of the main cluster varies by a factor of 3: $7 \times 10^{14}$ to $ 2 \times 10^{15} {\rm M_{\odot}}$, and the mass of the smaller cluster even more: $1 - 5 \times 10^{14} {\rm M_{\odot}}$.  The relative velocity at 2.5 Mpc in the different simulations ranges from $1.5 - 4.2 \times 10^{3}$ km/s.   This large disparity in inferred initial conditions demonstrates that the observations are in fact highly constraining vis-a-vis the initial conditions;  conversely, rough agreement between simulation and observations is insufficient to deduce the initial conditions and accurate modeling is needed.

Comparing other properties of the Bullet Cluster initial conditions inferred by SF14, we find that the impact parameter of the Bullet Cluster merger is compatible with the range found in analogous systems in simulated $\rm \Lambda CDM$ cosmology.  The concentration and shape of the less-massive initial cluster are also unremarkable, given its mass.   Available $\rm \Lambda CDM$ simulations do not have enough clusters in the extremely high mass range of the main cluster, $\rm \approx 2 \times 10^{15} M_\odot $, to permit a critical test of $\rm \Lambda CDM$ predictions for its halo concentration (1.3) and the ratio of the shortest to longest axes ($\rm \approx 0.35$), although they are not obviously incompatible with extrapolations from lower mass systems.  

We conclude that the initial conditions of the Bullet Cluster are compatible within uncertainties, with the range expected to occur in a $\rm \Lambda CDM$ cosmology.

\acknowledgments
Thanks to Jeremy Tinker for valuable discussions and advice in several areas, and to Scott Randall and Will Dawson for helpful consultations.  Thanks also to the anonymous referee for pointing out a number of items requiring clarification and making suggestions for improvements.  This work has been supported in part by grants NNX08AG70G, NSF PHY-1212538, NSF PHY-0900631 and NSF PHY-0970075.

\bibliographystyle{JHEP}
\bibliography{bullet}

\providecommand{\href}[2]{#2}\begingroup\raggedright\begin{thebibliography}{10}

\bibitem{Milgrom_2008}
M.~{Milgrom}, {\it {The MOND paradigm}},  {\em ArXiv e-prints} (Jan., 2008)
  [\href{http://xxx.lanl.gov/abs/0801.3133}{{\tt arXiv:0801.3133}}].

\bibitem{Moffat}
J.~R. {Brownstein} and J.~W. {Moffat}, {\it {The Bullet Cluster 1E0657-558
  evidence shows modified gravity in the absence of dark matter}},  {\em
  \mnras} {\bf 382} (Nov., 2007) 29--47,
  [\href{http://xxx.lanl.gov/abs/astro-ph/}{{\tt astro-ph/}}].

\bibitem{Lee_Komatsu}
J.~{Lee} and E.~{Komatsu}, {\it {Bullet Cluster: A Challenge to {$\Lambda$}CDM
  Cosmology}},  {\em \apj} {\bf 718} (July, 2010) 60--65,
  [\href{http://xxx.lanl.gov/abs/1003.0939}{{\tt arXiv:1003.0939}}].

\bibitem{milo2007}
M.~{Milosavljevi{\'c}}, J.~{Koda}, D.~{Nagai}, E.~{Nakar}, and P.~R. {Shapiro},
  {\it {The Cluster-Merger Shock in 1E 0657-56: Faster than a Speeding
  Bullet?}},  {\em \apjl} {\bf 661} (June, 2007) L131--L134,
  [\href{http://xxx.lanl.gov/abs/astro-ph/}{{\tt astro-ph/}}].

\bibitem{sf07}
V.~{Springel} and G.~R. {Farrar}, {\it {The speed of the `bullet' in the
  merging galaxy cluster 1E0657-56}},  {\em \mnras} {\bf 380} (Sept., 2007)
  911--925, [\href{http://xxx.lanl.gov/abs/astro-ph/}{{\tt astro-ph/}}].

\bibitem{Mastropietro}
C.~{Mastropietro} and A.~{Burkert}, {\it {Simulating the Bullet Cluster}},
  {\em \mnras} {\bf 389} (Sept., 2008) 967--988,
  [\href{http://xxx.lanl.gov/abs/0711.0967}{{\tt arXiv:0711.0967}}].

\bibitem{LageFarrar2014}
C.~{Lage} and G.~{Farrar}, {\it {Constrained Simulation of the Bullet
  Cluster}},  {\em \apj} {\bf 787} (June, 2014) 144,
  [\href{http://xxx.lanl.gov/abs/1312.0959}{{\tt arXiv:1312.0959}}].

\bibitem{Randall_2008}
S.~W. {Randall}, M.~{Markevitch}, D.~{Clowe}, A.~H. {Gonzalez}, and
  M.~{Brada{\v c}}, {\it {Constraints on the Self-Interaction Cross Section of
  Dark Matter from Numerical Simulations of the Merging Galaxy Cluster 1E
  0657-56}},  {\em \apj} {\bf 679} (June, 2008) 1173--1180,
  [\href{http://xxx.lanl.gov/abs/0704.0261}{{\tt arXiv:0704.0261}}].

\bibitem{Dawson2013}
W.~A. {Dawson}, {\it {The Dynamics of Merging Clusters: A Monte Carlo Solution
  Applied to the Bullet and Musket Ball Clusters}},  {\em \apj} {\bf 772}
  (Aug., 2013) 131, [\href{http://xxx.lanl.gov/abs/1210.0014}{{\tt
  arXiv:1210.0014}}].

\bibitem{Horizon_Paper}
J.~{Kim}, C.~{Park}, J.~R. {Gott}, III, and J.~{Dubinski}, {\it {The Horizon
  Run N-Body Simulation: Baryon Acoustic Oscillations and Topology of
  Large-scale Structure of the Universe}},  {\em \apj} {\bf 701} (Aug., 2009)
  1547--1559, [\href{http://xxx.lanl.gov/abs/0812.1392}{{\tt
  arXiv:0812.1392}}].

\bibitem{Comerford}
J.~M. {Comerford} and P.~{Natarajan}, {\it {The observed concentration-mass
  relation for galaxy clusters}},  {\em \mnras} {\bf 379} (July, 2007)
  190--200, [\href{http://xxx.lanl.gov/abs/astro-ph/}{{\tt astro-ph/}}].

\bibitem{Okabe2010}
N.~{Okabe}, M.~{Takada}, K.~{Umetsu}, T.~{Futamase}, and G.~P. {Smith}, {\it
  {LoCuSS: Subaru Weak Lensing Study of 30 Galaxy Clusters}},  {\em \pasj} {\bf
  62} (June, 2010) 811--, [\href{http://xxx.lanl.gov/abs/0903.1103}{{\tt
  arXiv:0903.1103}}].

\bibitem{bailin2005}
J.~{Bailin} and M.~{Steinmetz}, {\it {Internal and External Alignment of the
  Shapes and Angular Momenta of {$\Lambda$}CDM Halos}},  {\em \apj} {\bf 627}
  (July, 2005) 647--665, [\href{http://xxx.lanl.gov/abs/astro-ph/}{{\tt
  astro-ph/}}].

\bibitem{Schneider2012}
M.~D. {Schneider}, C.~S. {Frenk}, and S.~{Cole}, {\it {The shapes and
  alignments of dark matter halos}},  {\em \jcap} {\bf 5} (May, 2012) 30,
  [\href{http://xxx.lanl.gov/abs/1111.5616}{{\tt arXiv:1111.5616}}].

\bibitem{markevitch05}
M.~Markevitch, {\it Chandra observation of the most interesting cluster in the
  universe},  {\em astro-ph/0511345}.

\bibitem{fr07}
G.~R. {Farrar} and R.~A. {Rosen}, {\it {A New Force in the Dark Sector?}},
  {\em Physical Review Letters} {\bf 98} (Apr., 2007) 171302--+,
  [\href{http://xxx.lanl.gov/abs/astro-ph/}{{\tt astro-ph/}}].

\bibitem{hwRareBullet06}
E.~Hayashi and S.~White, {\it How rare is the bullet cluster?},  {\em
  astro-ph/0604443} (2006).

\end{thebibliography}\endgroup

\end{document}